\UseRawInputEncoding
\documentclass[twocolumn,prc,superscriptaddress,unsortedaddress,preprintnumbers,amsmath,amssymb]{revtex4}
\usepackage{tipa}
\usepackage{float}
\usepackage{color}
\usepackage[colorlinks,
			linkcolor=blue,
			anchorcolor=blue,
			citecolor=blue]{hyperref}
\usepackage{amsmath}
\definecolor{gray}{RGB}{220,220,220}

\usepackage{graphicx}
\usepackage{dcolumn}
\usepackage{bm}

\def\beq{\begin{equation}}
	\def\eeq{\end{equation}}
\def\bea{\begin{eqnarray}}
	\def\eea{\end{eqnarray}}

\def\fun#1#2{\lower3.6pt\vbox{\baselineskip0pt\lineskip.9pt
		\ialign{$\mathsurround=0pt#1\hfil##\hfil$\crcr#2\crcr$\sim$\crcr}}}
\preprint{}
\begin{document}
    \title{Phase diagrams of BCS-BEC crossover in asymmetric nuclear matter}
	\author{K. D. Duan}
        \affiliation{Institute of Modern Physics, Chinese Academy of Sciences, Lanzhou 730000, China}
        \affiliation{School of Nuclear Science and Technology, University of Chinese Academy of Sciences, Beijing 100049, China}
        \author{X. L. Shang}\email[]{shangxinle@impcas.ac.cn}
        \affiliation{Institute of Modern Physics, Chinese Academy of Sciences, Lanzhou 730000, China}
        \affiliation{School of Nuclear Science and Technology, University of Chinese Academy of Sciences, Beijing 100049, China}
        
	\begin{abstract}
    The phase structure of the BCS-BEC crossover for neutron-proton superfluid in asymmetric nuclear matter is systematically investigated, with particular focus on the roles of the angle-dependent gap (ADG) and the Fulde-Ferrell-Larkin-Ovchinnikov (FFLO) states. Phase diagrams in the $T-\alpha$, $\alpha-\rho$, and $T-\rho$ planes are constructed using both angle-averaged and angle-dependent gap treatments, enabling a unified analysis of the interplay between the FFLO pairing, ADG, and normal-superfluid phase separation (PS). The results confirm that the crossover is primarily density-driven. In the weak-coupling BCS regime, isospin asymmetry suppresses the stability of the homogeneous superfluid phase and drives the system toward PS, while the FFLO and ADG mechanisms partially alleviate this suppression. Although the ADG itself does not extend the asymmetry window for superfluidity, in combination with the FFLO state it enlarges the asymmetry range over which superfluidity survives and significantly reduces PS. At high density, these combined effects can nearly eliminate PS. However, as density decreases, ADG-induced suppression of PS is progressively weakened, due to both the reduced destructive effect of isospin asymmetry and the decreasing $D$-wave fraction in the $^3SD_1$ channel. In general, the system evolves smoothly from a $D$-wave–dominated superfluid at high density to a $S$-wave superfluid at low density, with a corresponding weakening of ADG effects. Furthermore, the ADG lifts the orientational degeneracy of the FFLO state, resulting in two distinct FFLO-ADG phases separated by a first-order transition. In contrast, in the BEC regime, the FFLO and ADG states vanish, while the PS persists, leading to an inhomogeneous mixed phase at low temperatures and large asymmetries, where the superfluid component forms a BEC of deuterons. 
	\end{abstract}
    
    \maketitle
    
    \section{Introduction}
    Within the Bardeen-Cooper-Schrieffer (BCS) theory, fermionic superfluids originate from the formation of boson-like Cooper pairs, where two fermions form a loosely bound state by attractive interaction. With increasing pairing strength, these pairs undergo a smooth crossover to a Bose-Einstein condensate (BEC) of tightly bound dimers \cite{Eagles1969,Leggett1980,Nozieres1985}. 
    This BCS-BEC crossover phenomenon has been experimentally verified in ultracold Fermi gases by tuning the interaction strength through Feshbach resonances \cite{Jochim2003,Zwierlein2003,Greiner2003,Zwierlein2005}. A similar BCS-BEC crossover is also expected to occur in nuclear matter \cite{Alm1993,Baldo1995,Stein1995,Andrenacci1999,Lombardo2001}, where neutron-proton ($np$) pairs smoothly evolve from $np$ Cooper pairs at higher densities to a gas of Bose-condensed deuterons at extremely low nucleon densities. This mechanism may play a significant role in various phenomena such as deuteron formation in heavy-ion collisions \cite{Baldo1995} and supernovae \cite{Heckel2009,Typel2010,Stein1995} at intermediate energies.  In the weak coupling BCS regime, $np$ pairing provides crucial insight into the structure of $N \simeq Z$ finite nuclei \cite{Wang2024, Goodman1999,Ropke2000,Frauendorf2014}, offering valuable information on the underlying nuclear interaction \cite{shang2023}. In addition, $np$ pairing properties may be a key ingredient underlying the cooling and rotational dynamics in neutron star models, which permit pion or kaon condensation \cite{Brown1994}.

    Since the pairing property depends crucially on the overlap between the wave functions of the two paired particles near the Fermi surface, $np$ pairing is significantly hindered in asymmetric nuclear matter due to population imbalances. Beyond this suppressing effect, such imbalances in a two-component superfluid system can further give rise to a variety of exotic quantum phenomena, including nonzero Cooper pair momentum \cite{Fulde1964,Larkin1964}, gapless excitations \cite{Sarma1963}, normal-superfluid phase separation (PS) \cite{Shin2006,Sheehy2006,Bedaque2003}, and superfluid-density instability \cite{Pao2006,Wu2003,He200612}. These phenomena have attracted considerable interest in both theoretical \cite{Sheehy2006,Pao200674,He2006Dec,wang2018,Shang2022} and experimental \cite{Zwierlein2006,Partridge2006,Shin2006} studies of analogous asymmetric two-component superfluid systems, such as ultracold atomic Fermi gases and neutral dense quark matter. The essential origin of these unconventional phenomena lies in the requirement to accommodate excess unpaired particles in the phase space near the Fermi surface, which substantially alters the quantum properties of the system. Compared to isotropic $S$-wave pairing, non-$S$-wave pairing channels provide suitable phase spaces to settle the excess unpaired neutrons \cite{Wang2015}. As a result, the angular dependence of the pairing gap inherent to non-$S$-wave pairing can naturally mitigate the detrimental influence of asymmetry on $np$ pairing \cite{Shang2013}. 

    Due to the tensor force, the dominant component of the $np$ pairing below the nuclear saturation density \cite{Alm1993,Sedrakian2000,Lombardo2001} corresponds to the attractive isospin-singlet $^3SD_1$ channel, which involves a mixing of the $S$-wave and the $D$-wave components. This hybrid structure of the pairing gap leads to a phase diagram \cite{Duan2025} that is distinct from those of either pure $S$-wave \cite{He2006Dec} or $D$-wave \cite{Shang2022} pairing in the weak-coupling BCS regime. In particular, the angular dependence of the pairing gap originating from the anisotropic $D$-wave can eliminate the normal-superfluid PS in the low-asymmetry region \cite{Duan2025}, which is markedly different from the asymmetry range associated with the pair momentum in the Fulde-Ferrell-Larkin-Ovchinnikov (FFLO) state \cite{Fulde1964,Larkin1964}. However, the role of this angular dependence in the strong-coupling regime remains unclear. Specifically, its influence on the $np$ BCS-BEC crossover has never been investigated. The main objective of this work is therefore to incorporate the angle-dependent pairing gap in the $^3SD_1$ channel into a model of isospin-asymmetric nuclear matter and construct the corresponding phase diagram over wide ranges of density, temperature and isospin asymmetry, including non-BCS pairing states. Particular attention is devoted to the impact of the angular dependence on the phase structure.  

    The paper is organized as follows. Section \ref{formalism} outlines the theoretical framework, including the evolution from the BCS equation toward the BEC limit, the thermodynamic stability conditions, and the stability criterion of homogeneous states against phase separation. In Sec. \ref{results and discussions}, we present the phase diagrams for the BCS-BEC crossover, considering both conventional (zero-momentum) and FFLO (finite-momentum) pairing states. Finally, a summary and outlook are given in Sec. \ref{summary and outlook}. Throughout this paper, we adopt natural units, $c = \hbar = k_B = 1$.

    \section{Formalism} \label{formalism}
    In this section, we briefly review the treatment of both the weak coupling BCS regime and the strong coupling BEC regime within the BCS framework \cite{Baldo1995,Lombardo2001,Lombardo2001size}. We start with the finite-temperature gap equation at fixed number densities, which reads:
    \begin{align}\label{gap}
    \Delta(\mathbf{k})=\
    &\sum_{\mathbf{k}^{\prime}}V(\mathbf{k},\mathbf{k}^{\prime})\frac{\Delta(\mathbf{k}^{\prime})}{2 \sqrt{\xi_{\mathbf{k}^{\prime}}^{2}+D^{2}(\mathbf{k}^{\prime})}} \nonumber\\[2mm]
    &\times\left[1-f\left(E_{\mathbf{k}^{\prime}}^{+}\right)-f\left(E_{\mathbf{k}^{\prime}}^{-}\right)\right],
    \end{align}
    together with the density constraints:
    \begin{align}\label{density}
        \rho_{p/n} =\sum_{\mathbf{k}}&n_{p/n}(\mathbf{k}), \nonumber\\
        =\sum_{\mathbf{k}}&\Big\{\frac{1}{2}\Big[1+\frac{\xi_{\mathbf{k}}}{\sqrt{\xi_{\mathbf{k}}^{2}+D^{2}(\mathbf{k})}}\Big]f(E_{\mathbf{k}}^{\pm}) \nonumber\\
        &+\frac{1}{2}\Big[1-\frac{\xi_{\mathbf{k}}}{\sqrt{\xi_{\mathbf{k}}^{2}+D^{2}(\mathbf{k})}}\Big]\Big[1-f(E_{\mathbf{k}}^{\mp})\Big]\Big\}.
    \end{align}
    Here, $V(\mathbf{k},\mathbf{k}^{\prime})$ denotes the nucleon-nucleon ($NN$) interaction in the relevant channel. 
    The quasiparticle excitation spectrum is given by $E_{\mathbf{k}}^{\pm}=\sqrt{\xi_{\mathbf{k}}^{2}+D^{2}(\mathbf{k})}\pm\delta\varepsilon_{\mathbf{k}}$ with $\xi_{\mathbf{k}}=\frac{1}{2}(\varepsilon_p+\varepsilon_n)$ and $\delta\varepsilon_{\mathbf{k}}=\frac{1}{2}(\varepsilon_p-\varepsilon_n)$.
    The pairing gap $\Delta(\mathbf{k})$ represents the gap matrix in the spin space and possesses the unitarity property $\Delta(\mathbf{k})\Delta^{\dag}(\mathbf{k})=ID^{2}(\mathbf{k})$ in the pure $^3SD_1$ channel, where $I$ represents the identity matrix and $D^2(\mathbf{k})$ is a scalar quantity.
    The single-particle energy for neutron/proton is defined as $\varepsilon_{n/p}=\frac{(\mathbf{q}\pm\mathbf{k})^{2}}{2m}-\mu_{n/p}$, with $\mu_{n/p}$ the chemical potentials obtained self-consistently from Eqs.~(\ref{gap}) and (\ref{density}). The vector $2\mathbf{q}$ represents the nonzero center-of-mass momentum of a Cooper pair in the FFLO state.
    The finite-temperature Fermi-Dirac distribution is $f(x)=\big[1+e^{x/T}\big]^{-1}$ with $T$ the temperature.

    The pairing probability, which describes the dynamics of the Cooper pairs in the condensate, is defined as:
    \begin{align}\label{psi}
    \kappa_\mathbf{k} &= \frac{\Delta_\mathbf{k}}{2E_\mathbf{k}} \left[1-f(E_{\mathbf{k}}^{+})-f(E_{\mathbf{k}}^{-})\right], 
    \end{align}
    where $\xi_{\Delta}=\sqrt{\xi_{\mathbf{k}}^{2}+D^{2}(\mathbf{k})}$. The total number-density distribution satisfies:
     \begin{align} \label{denc}
    1-n_{p}-n_{n} &= \frac{\xi_{\mathbf{k}}}{\xi_{\Delta}} \left[1-f(E_{\mathbf{k}}^{+})-f(E_{\mathbf{k}}^{-})\right].
    \end{align}
    Using Eqs. (\ref{psi}) and (\ref{denc}) the gap equation can be rewritten in the form:
    \begin{equation}
    \frac{\mathbf{k}^2}{m} \kappa_\mathbf{k} + (1 - n_p-n_n) \sum_{\mathbf{k}^{\prime}} V(\mathbf{k},\mathbf{k}^{\prime}) \kappa_{\mathbf{k}^{\prime}} = 2\mu \kappa_\mathbf{k}.
    \end{equation}
    In the low-density limit ($n_{p/n} \to 0$), this equation reduces to the Schr\"{o}dinger equation for the deuteron bound state \cite{Baldo1995,Lombardo2001}. In this regime, the pairing probability $\kappa_\mathbf{k}$ reduces to the deuteron wave function, and the chemical potential $2\mu=\mu_n+\mu_p$ approaches the corresponding bound-state energy. The crossover from the BCS to the BEC regime can be identified through the sign of the chemical potential $\mu$ \cite{Baldo1995}.  A positive value ($\mu>0$) implies the presence of a well-defined Fermi surface, manifested by an inflection point in the occupation-number distribution, whereas a negative value ($\mu<0$) signals the disappearance of the Fermi sea and the emergence of tightly bound bosonic pairs \cite{Strinati2018}.

    Owing to the tensor force, the hybrid $^3SD_1$ pairing structure gives rise to an intrinsic anisotropic pairing gap. In early studies, this angular dependence was often eliminated using an angle-averaging procedure for simplicity \cite{Baldo1995,Ropke2000,Lombardo2001,Sedrakian2001}, in which $D^{2}(\mathbf{k})$ is replaced by $\frac{1}{4\pi}\int d \Omega_{\mathbf{k}}D^{2}(\mathbf{k})$. However, the angular dependence becomes particularly important in asymmetric nuclear matter, as it can reduce the suppression caused by isospin asymmetry \cite{Shang2013,Duan2025}. For clarity of comparison, we refer to the gap obtained under the angle-averaging procedure as the angle-averaged gap (AAG), while the gap preserving its angular dependence is termed the angle-dependent gap (ADG), following Ref. \cite{Shang2013}. Similarly as proposed in Ref. \cite{Shang2013}, we adopt the axial-symmetric gap ansatz, for which $D^2(\mathbf{k})$ takes the form
    \begin{align}\label{function-ADG spectrum}
        &D^{2}(\mathbf{k}) = \Delta_{0}^{2}(k) + \sqrt{2} \Delta_{0}(k) \Delta_{2}(k)\left(3 \cos^{2} \theta - 1\right)  \nonumber\\[2mm]
        &\quad \quad \quad \quad + \Delta_{2}^{2}(k)\left(\frac{3 \cos^{2} \theta + 1}{2}\right),
    \end{align}
    where $(\theta, \phi)$ denotes the orientation of $\mathbf{k}$ in spherical coordinates. With this ansatz, the gap equation becomes
    \begin{align}\label{function-ADG}
	\left(\begin{array}{l}
		\Delta_{0} \\
		\Delta_{2}
	       \end{array}\right)(k) & = \frac{-1}{\pi} \int d k^{\prime}k^{\prime 2}\left(\begin{array}{ll}
		V^{00} & V^{02} \\
		V^{20} & V^{22}
	\end{array}\right)\left(k, k^{\prime}\right) \nonumber\\[2mm]
		&\times 
	\int \frac{d \Omega_{\mathbf{k}^{\prime}}}{4\pi} \frac{1-f\left(E_{k^{\prime}}^{+}\right)-f\left(E_{k^{\prime}}^{-}\right)}{\sqrt{\xi_{\mathbf{k}^{\prime}}^{2}+D^{2}\left(\mathbf{k}^{\prime}\right)}}\nonumber\\[2mm]
        &\times\left(\begin{array}{ll}
            \ \ \ \ \ \ 1 & \ \ \frac{3 \cos^{2} \theta - 1}{\sqrt{2}} \\
            \frac{3 \cos^{2} \theta - 1}{\sqrt{2}} & \ \ \frac{3 \cos^{2} \theta + 1}{2}
        \end{array}\right)
	\left(\begin{array}{c}
		\Delta_{0} \\
		\Delta_{2}
	\end{array}\right)\left(k^{\prime}\right),
    \end{align}
    where $V^{00}$, $V^{02}$, $V^{20}$, and $V^{22}$ are the matrix elements of the $NN$ interaction in the $^3SD_1$ channel. The off-diagonal elements $V^{02}$ ($V^{20}$), originating from the tensor force, couple the $S$-wave component $\Delta_{0}$ and the $D$-wave component $\Delta_{2}$. Unless otherwise specified, the ‘pairing gap’ refers to $\Delta(k)=\sqrt{\Delta_0(k)^2+\Delta_2(k)^2}$ for a consistent comparison with other references. 

    When the pairing gap becomes anisotropic, the quasiparticle spectrum no longer degenerates with respect to the orientation of the Cooper pair momentum. To account for this in the FFLO state, particular attention need be paid to the symmetric and antisymmetric parts of the spectrum:
    \begin{align}\label{function-spectrum}
	\xi_{\mathbf{k}} & =\frac{\mathbf{k}^{2}}{2 m}+\frac{\mathbf{q}^{2}}{2 m}-\mu, \nonumber\\
	\delta \varepsilon & =h-\frac{\mathbf{k} \cdot \mathbf{q}}{m}=h-\frac{k q}{m} \cos (\widehat{\mathbf{k q}}) \nonumber\\
	& =h-\frac{k q}{m}\left[\sin \theta_{0} \sin \theta \cos \left(\phi-\phi_{0}\right)+\cos \theta_{0} \cos \theta\right],
    \end{align}
    where $\mu=(\mu_n+\mu_p)/2$ is the average chemical potential and $h=(\mu_n-\mu_p)/2$ represents the Fermi-surface mismatch. The spherical angles $(\theta_0,\phi_0)$ specify the orientation of $\mathbf{q}$ in spherical coordinates. The term $\frac{\mathbf{k} \cdot \mathbf{q}}{m}$, originating from the Cooper pair momentum, reduces the suppression induced by the Fermi-surface mismatch $ℎ$ along certain directions. Without loss of generality, we choose a coordinate system in which the direction of $\mathbf{q}$ is fixed at $(\theta_0, 0)$, thus eliminating the dependence on $\phi_{0}$. In this scenario, only $\theta_0$ is retained to specify the direction of the Cooper pair momentum. Consequently, once the angular dependence of the pairing gap is taken into account, two parameters are introduced, the magnitude $q$ and the angle $\theta_0$, to describe the Cooper pair momentum. These parameters must be determined by minimizing the total energy of the system.

    For asymmetric nuclear matter at finite temperature $T$, the entropy $S$ and the internal energy $U$ of the superfluid state can be expressed as
    \begin{align}\label{function-entropy}
	S =& -2\sum_{\mathbf{k}}\Big\{f(E_{\mathbf{k}}^+)\ln f(E_{\mathbf{k}}^+) + \big[1-f(E_{\mathbf{k}}^+)\big]\ln\big[1-f(E_{\mathbf{k}}^+)\big] \nonumber\\[2mm]
	&+ f(E_{\mathbf{k}}^-)\ln f(E_{\mathbf{k}}^-) + \big[1-f(E_{\mathbf{k}}^-)\big]\ln\big[1-f(E_{\mathbf{k}}^-)\big]\Big\},
    \end{align}
    \begin{align}\label{function-internal energy}
	U = \sum_{\mathbf{k}}\Big[\varepsilon_n n_n(\mathbf{k}) +\varepsilon_p n_p(\mathbf{k})\Big] - \sum_{\mathbf{k},\mathbf{k}^{\prime}}V(\mathbf{k},\mathbf{k}^{\prime})\nu^{\dagger}(\mathbf{k})\nu(\mathbf{k}^{\prime}).
    \end{align}
    The essential thermodynamic quantity characterizing the system is the free energy
    \begin{equation}\label{function-free energy}
        \mathcal{F} = U  - TS  + \mu\rho + h \delta\rho,
    \end{equation}
    where $\rho=\rho_n + \rho_p$ is the total number density and $\delta\rho=\rho_n - \rho_p$ is the population imbalance. The isospin asymmetry is defined as $\alpha=\delta\rho/\rho$.

    The thermodynamically stable solution should minimize the free energy difference between the superfluid and normal phase, $\delta\mathcal{F}=\mathcal{F}_{S}-\mathcal{F}_{N}$, where $\mathcal{F}_{S}$ can be obtained from Eq. (\ref{function-free energy}) in the limit $\Delta\rightarrow 0$. The magnitude $q$ and orientation $\theta_0$ of the Cooper pair momentum are determined by
    \begin{equation}\label{function-q}
	\frac{\mathcal{D}\mathcal{F}}{\mathcal{D}q} = 0, \ \ \frac{\mathcal{D}\mathcal{F}}{\mathcal{D}\theta_{0}} = 0,
    \end{equation}
    with the full differentials defined as
    \begin{align}
		 \frac{\mathcal{D}}{\mathcal{D}q}&=\frac{\partial}{\partial q}+\frac{\partial\mu}{\partial q}\frac{\partial}{\partial \mu}+\frac{\partial h}{\partial q}\frac{\partial}{\partial h}, \nonumber \\
        \frac{\mathcal{D}}{\mathcal{D}\theta_{0}}&=\frac{\partial}{\partial \theta_{0}}+\frac{\partial\mu}{\partial \theta_{0}}\frac{\partial}{\partial \mu}+\frac{\partial h}{\partial \theta_{0}}\frac{\partial}{\partial h}.
	  \end{align}
    These full differentials arise from the fact that both $q$ and $\theta_0$ depend implicitly on $\mu$ and $h$ through the density constraints (\ref{density}). Using these relations one finds $\mathcal{D}\mathcal{F}/\mathcal{D}q=\partial\Omega/\partial q$ and $\mathcal{D}\mathcal{F}/\mathcal{D}\theta_{0}=\partial\Omega/\partial \theta_{0}$. It should be emphasized that Eq. (\ref{function-q}) must be solved self-consistently together with the gap equation (\ref{function-ADG}) and the density constraints (\ref{density}).

        \subsection{Superfluid density}
        The superfluid density, which must remain positive to ensure the thermodynamic stability of the system \cite{Pao2006,Pao200674}, characterizes the tendency for a finite Cooper pair momentum to emerge. It can be evaluated from the linear response of the free energy to a tiny superfluid velocity $\mathbf{v}_{s}$, and is defined as
        \begin{equation}\label{rhos}
            \rho_s = \frac{\mathcal{D}^2\mathcal{F}}{\mathcal{D}\mathbf{v}_{s}^2}\Big|_{\mathbf{v}_{s}=0}=m^2\frac{\mathcal{D}^2\mathcal{F}}{\mathcal{D}\mathbf{q}^2}\Big|_{\mathbf{q}=0}=\frac{\partial^2\Omega}{\partial\mathbf{q}^2}\Big|_{\mathbf{q}=0},
        \end{equation}
        for the conventional BCS state in the absence of Cooper pair momentum. A positive superfluid density corresponds to a positive-definite matrix $\frac{\mathcal{D}^2\mathcal{F}}{\mathcal{D}\mathbf{q}}$.
        Due to the axial symmetry of the ADG state, the anisotropic superfluid-density tensor can be decomposed into longitudinal and transverse components under the residual $O(2)$ symmetry, defined as
        \begin{equation}\label{function-density L&T}
            \rho_L = m\rho + \int \frac{k^4 \mathrm{d}k}{4\pi^2} \int_{-1}^{1} \cos^2 \theta  \mathrm{~d}\cos\theta (f'(\tilde{E}_{\mathbf{k}}^+) + f'(\tilde{E}_{\mathbf{k}}^-))\nonumber,
        \end{equation}
        \begin{equation}
            \rho_T = m\rho + \int \frac{k^4 \mathrm{d}k}{8\pi^2} \int_{-1}^{1} \sin^2 \theta  \mathrm{~d}\cos\theta (f'(\tilde{E}_{\mathbf{k}}^+) + f'(\tilde{E}_{\mathbf{k}}^-)),
        \end{equation}    
        where $\tilde{E}_{\mathbf{k}}^{\pm} = \left.E_{\mathbf{k}}^{\pm} \right|_{q=0}$. Once the pairing gap becomes isotropic, the longitudinal and transverse components coincide, yielding $\rho_L = \rho_T= \rho_s$. In contrast, for the ADG scenario, the discrepancy between $\rho_L$ and $\rho_T$ indicates that the tendency for the Cooper pair momentum to emerge is direction-dependent, suggesting that multiple possible orientations of the FFLO Cooper pair momentum may emerge.

        \subsection{Stability condition against phase separation in the homogeneous superfluid state}
        Note that the homogeneous superfluid phase may become unstable and evolve into an inhomogeneous mixed phase composed of coexisting normal and superfluid components \cite{Bedaque2003,Viverit2000,Cohen2005}, in which the suppression of $np$ pairing induced by asymmetry can also be mitigated. The stability of the homogeneous superfluid phase can be evaluated by examining the response of the total free energy $F=\int d^3\mathbf{r}\mathcal{F}[\rho_\tau(\mathbf{r})]$ (here $\tau=n, p$) to small fluctuations $\delta\rho_\tau(\mathbf{r})$ of the local densities imposed on top of the homogeneous density distribution. Since the first-order variation vanishes, the stability condition against phase separation is determined by the second-order variation $\delta^2F$. A homogeneous superfluid phase is stable only if the matrix $\partial^{2} \mathcal{F} / \partial \rho_{\tau} \partial \rho_{\tau'} = \partial\mu_{\tau'} / \partial\rho_\tau$ is positive definite. By transforming the density variables to $\rho$ and $\delta\rho$, this condition becomes equivalent to demanding the positive definiteness of the matrix
        \begin{align}\label{equ:M22}
            \mathcal{M}=\left(\begin{array}{cc}
        	\frac{\partial\mu}{\partial\rho} & \frac{\partial\mu}{\partial\delta\rho} \\
        	\frac{\partial h}{\partial\rho} & \frac{\partial h}{\partial\delta\rho}
            \end{array}\right).
        \end{align}
        The eigenvalues of $\mathcal{M}$
        \begin{equation}\label{equ:eigenvalue}
            \lambda_{\pm}=\frac{\operatorname{Tr}\mathcal{M} \pm \sqrt{\operatorname{Tr}\mathcal{M}^{2}-4 \operatorname{det}\left(\mathcal{M}\right)}}{2},
        \end{equation}
        must both be positive for the stable homogeneous superfluid state. Here, $\operatorname{Tr}\mathcal{M}=\partial\mu / \partial\rho+ \partial h / \partial\delta\rho$. The eigenvalue $\lambda_+$, which governs the stability of the total density distribution, is always positive in asymmetric nuclear matter. In contrast, $\lambda_-$ is roughly associated with the isospin asymmetry distribution and therefore determines whether the homogeneous superfluid phase is stable or becomes unstable toward phase separation. 
        
    \section{Results and Discussions}\label{results and discussions}
    The numerical calculations in this work focus on constructing the phase diagram of the BCS-BEC crossover in neutron-proton superfluidity for asymmetric nuclear matter. To simplify the calculations, several approximations have been adopted. First, the competition between isospin-singlet and isospin-triplet pairing channels is not considered. Although both channels may coexist in the BCS regime \cite{Yan2021}, neither $nn$ nor $pp$ bound states can form in the BEC limit, and the $nn$ pairing closest to a BEC-like behavior is expected to appear only in the crossover region \cite{Sun2010}. Second, the bare $NN$ potential, specifically the Argonne $V18$ (Av18) potential, is employed as the pairing interaction. This treatment neglects the medium-induced screening effects, which, despite several studies \cite{Schulze1996,shen2005,cao2006,zhang2016}, remain an open issue for $np$ pairing. Third, a free single-particle (s.p.) spectrum is used to facilitate the calculation of the second derivative of the free energy. While incorporating self-energy effects would provide more realistic results \cite{Fan2019}, doing so introduces substantial computational challenges to evaluate the stability conditions. However, although the latter two approximations may modify the quantitative values of the pairing gap, the overall structure of the phase diagram is expected to remain qualitatively robust.

        \subsection{BCS-BEC crossover with $\mathbf{q}=0$}
        
\begin{figure}[h]
         	\centering
         	\includegraphics[width=1.0\linewidth]{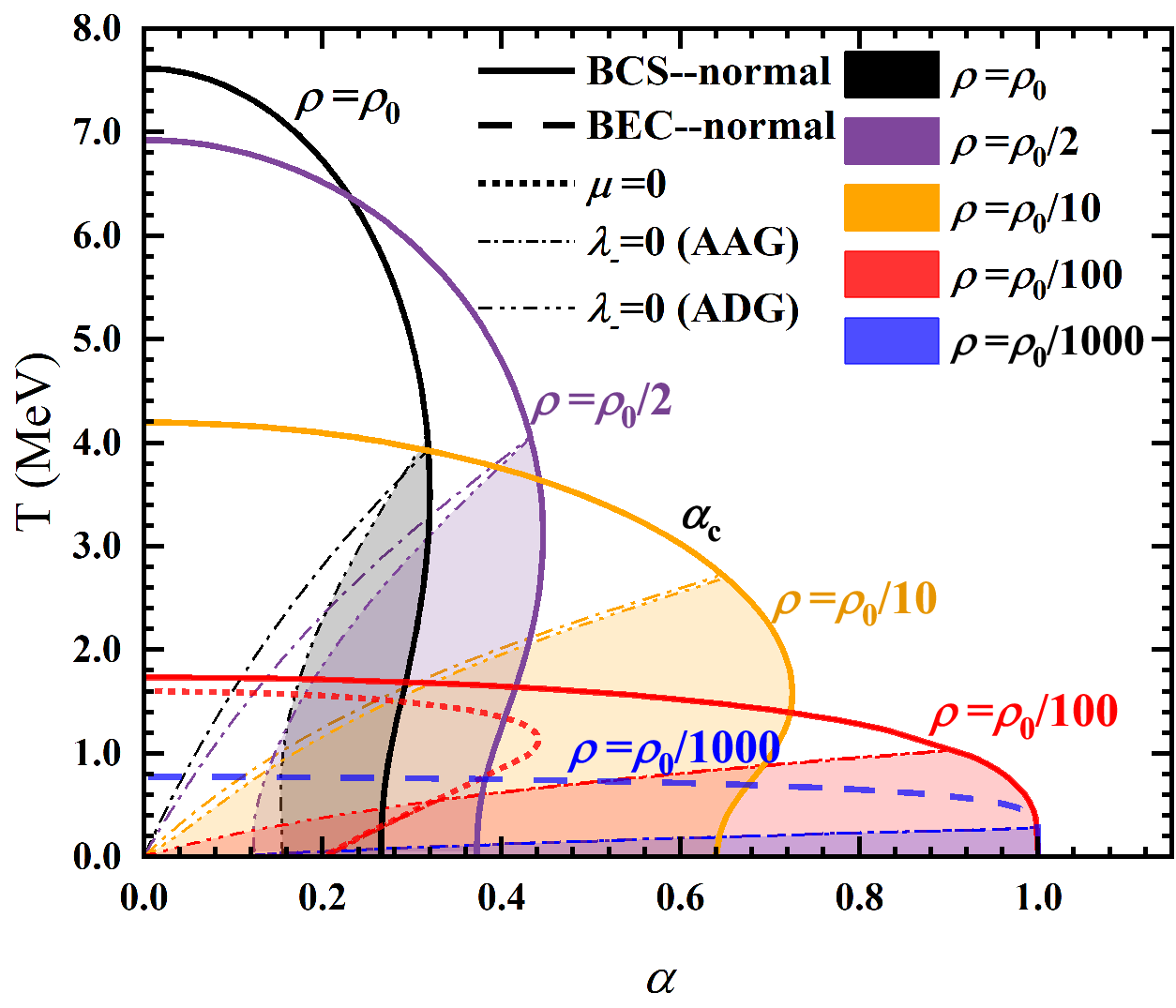}
         	\caption{(Color online)
            Phase diagram of the ADG and the AAG states in the $T-\alpha$ plane. The thick solid and dashed lines indicate the phase boundaries separating the BCS and BEC superfluid phases from the normal phase, respectively. The thick dotted line marks the condition $\mu=0$, signaling the onset of the BCS-BEC crossover. The thin dash-dotted and dash-dot-dotted lines denote the stability condition against PS for the AAG and ADG states, respectively. The shaded region highlights the normal-superfluid PS domain of the ADG state. The black, purple, orange, red, and blue colors correspond to densities $\rho=\rho_0$, $\rho=\rho_0/2$, $\rho=\rho_0/10$, $\rho=\rho_0/100$, $\rho=\rho_0/1000$, respectively, where $\rho_0=0.17~\text{fm}^{-3}$ is the saturation density.}
         	\label{fig:alpha-t}
        \end{figure}

        In nuclear matter, the effective coupling strength of the $np$ pairs increases as the density decreases; therefore, the density effectively serves as the parameter characterizing the coupling strength. Fig. \ref{fig:alpha-t} shows the phase diagrams in the $T-\alpha$ plane for several densities. At saturation density $\rho_0$, the ADG and AAG phases vanish at the same critical asymmetry $\alpha_c$. As seen in Fig. \ref{fig:alpha-t}, the phase boundaries corresponding to the second-order superfluid-normal transition for the ADG and AAG states coincide, consistent with Refs. \cite{Shang2013,Duan2025}. This indicates that the angular dependence of the pairing gap itself does not extend the asymmetry range over which the pairing gap can exist. However, it can eliminate normal-superfluid PS for lower asymmetries, as evidenced by the pronounced difference between the normal-superfluid PS conditions $\lambda_{-}$ for the ADG and AAG states. As the density decreases, the effective coupling strength increases while the suppressing effect induced by asymmetry becomes weaker, as indicated by the variation of $\alpha_c$ with density. In particular, $\alpha_c$ approaches unity for $\rho_0/1000$ at low temperature. In this BEC regime, the system consists of a BEC of deuterons and excess neutrons, and the perturbation from these excess neutrons to the deuteron bound states is negligible. Meanwhile, as shown in Fig. \ref{fig:alpha-t}, the mitigating effect of the angular dependence of the pairing gap against the asymmetry is progressively reduced with decreasing density. This behavior originates from the enhancement of the effective $np$ coupling strength at low densities, which weakens the destructive impact of asymmetry on $np$ pairing and, consequently, diminishes the additional alleviation provided by the ADG. As a result, the difference between the $\lambda_{-}$ values of the ADG and AAG states decreases and eventually becomes negligible.
        
        Additionally, a BCS-BEC crossover arises as the density decreases. At high density, only the BCS phase appears in the phase diagram, as exemplified by the case of $\rho_0$. At lower densities, the BEC phase emerges, first appearing at zero temperature in the symmetric case and subsequently extending to a larger region of the $T-\alpha$ plane. As shown in Fig. \ref{fig:alpha-t} for density $\rho_0/100$, the thick red dotted line denotes the BCS-BEC crossover ($\mu=0$): the region to its left corresponds to the BEC superfluid, whereas the region to its right remains in the BCS regime. At an even lower density of $\rho_0/1000$, only the BEC superfluid survives in the phase diagram. In this case, the homogeneous BEC phase and the PS-BEC phase are realized in the relatively high- and low-temperature regions, respectively. This behavior is consistent with that in the BCS regime, where thermal excitations can suppress PS induced by isospin asymmetry.

        Since the effective coupling strength of the $np$ pairing varies with density, the BCS-BEC crossover is primarily density-driven. The upper panels of Fig. \ref{fig:alpha-rho} show the phase diagrams of the AAG [Fig. \ref{fig:alpha-rho}(a)] and ADG [Fig. \ref{fig:alpha-rho}(b)] states in the $\alpha-\rho$ plane at temperature $0.1$ MeV, illustrating both the density-induced BCS-BEC crossover and the role of the angular dependence of the pairing gap. For the AAG state, the angle-averaging approximation reduces the pairing gap to an effective $S$-wave form, resulting in a phase diagram [Fig. \ref{fig:alpha-rho}(a)] consistent with that of $S$-wave pairing \cite{Chen2006}. The distinct signal of the crossover $\mu=0$, indicated by the thick magenta dotted line, separates the BCS regime at higher densities from the BEC regime at lower densities. With decreasing density, the system undergoes a smooth BCS-BEC crossover. During this process, the suppression of pairing due to isospin asymmetry gradually weakens, as reflected by the critical asymmetry $\alpha_c$, which increases from zero to unity and remains $\alpha_c=1$ throughout the BEC regime. The thin red dashed line marks the zero value of the superfluid density $\rho_s=0$. Since negative $\rho_s$ signals a potential instability toward the FFLO state, the confinement of the $\rho_s<0$ region to the BCS side indicates that the FFLO state is unlikely to emerge in the BEC regime. 

        Compared with the AAG case, the ADG state exhibits a substantially reduced normal-superfluid PS region, particularly near saturation density. This highlights the domain over which the angular dependence of the pairing gap significantly affects the PS. In this case, the superfluid density splits into transverse $\rho_T$ and longitudinal $\rho_L$ components, denoted by the thin red dashed and blue short-dashed lines, respectively. The distinct behaviors of $\rho_T$ and $\rho_L$ demonstrate that the angular dependence of the pairing gap is significant only in the weak-coupling BCS regime. 

        \subsection{Phase diagram including the FFLO state}
\begin{figure*}[t!]
            \centering
            \includegraphics[width=1.0\linewidth]{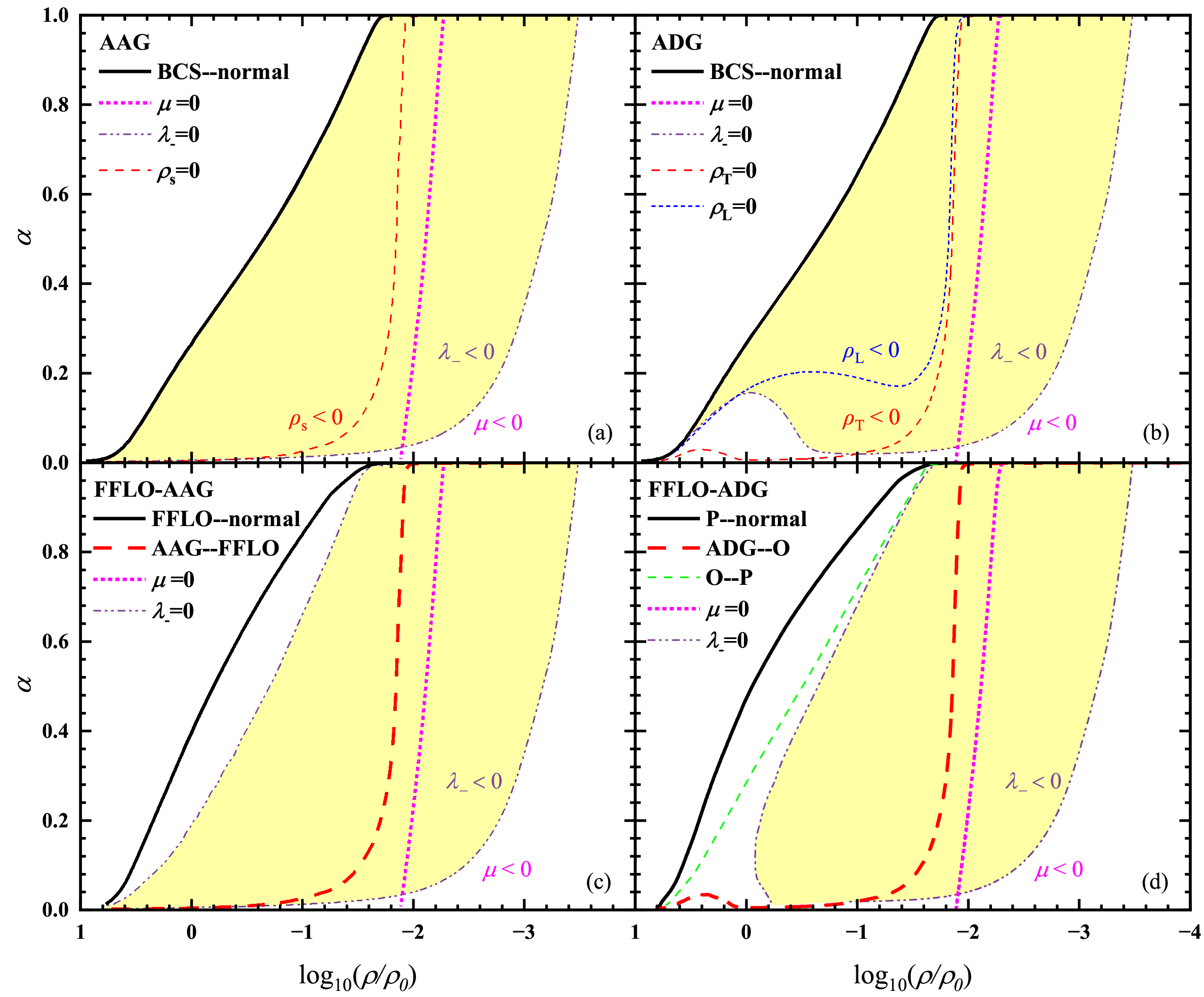}
            \caption{(Color online)
            Phase diagram of the AAG (a), ADG (b), FFLO-AAG (c) and FFLO-ADG (d) states in the $\alpha-\rho$ plane. The thick solid lines denote the phase boundary corresponding to the superfluid-normal transition. The thick dashed lines mark the transition from the FFLO state to the conventional BCS state. The thick dotted lines identify the condition $\mu=0$. The yellow shaded region corresponds to the domain of PS, while the thin dash-dot-dotted lines indicate the stability condition against PS in the homogeneous superfluid state. For the AAG state, the thin dashed line represents the zero value of the superfluid density. For the ADG state, the thin dashed and short-dashed lines correspond to the zero values of the transverse ($\rho_T$) and longitudinal ($\rho_L$) components of the superfluid density, respectively. For the FFLO-ADG state, the thin dashed line represents the phase transition from FFLO-ADG-O state to FFLO-ADG-P state (O-P).}
            \label{fig:alpha-rho}
        \end{figure*}

        As indicated by the superfluid density, the FFLO state with finite Cooper pair momentum is expected to appear in the BCS regime of the phase diagram. To investigate the specific characteristics of the BCS-BEC crossover in the presence of the FFLO state, the lower panels of Fig. \ref{fig:alpha-rho} display the phase diagrams in the $\alpha-\rho$ plane at temperature $0.1$ MeV, considering the angle-averaged gap [FFLO-AAG, Fig. \ref{fig:alpha-rho} (c)] and angle-dependent gap [FFLO-ADG, Fig. \ref{fig:alpha-rho} (d)]. The inclusion of a finite Cooper pair momentum in the FFLO-AAG state not only extends the range of isospin asymmetry over which the superfluid phase exists in the weak-coupling BCS regime, but also suppresses the normal-superfluid PS at large asymmetries, in line with the FFLO mechanism discussed in the literature \cite{Duan2025,Chen2006}. The emergence of the FFLO-AAG state coincides precisely with the boundary where the superfluid density of the AAG state approaches zero. In the strong-coupling BEC regime, the phase structure of the FFLO-AAG state completely overlaps with that of the AAG state, consistent with the phase diagram of an $S$-wave superfluid \cite{Chen2006}.

        When the angular dependence of the pairing gap is taken into account, the FFLO state becomes nondegenerate with respect to the orientation of the Cooper pair momentum. The transverse $\rho_T$ and longitudinal $\rho_L$ components of the superfluid density in ADG state suggest two distinct preferred directions, giving rise to two FFLO-ADG states: FFLO-ADG-O and the FFLO-ADG-P, corresponding to Cooper pair momentum directions $\theta_0=0$ and $\theta_0=\pi/2$, respectively \cite{Shang2015}. While intermediate orientations may appear in pure $D$-wave pairing \cite{Zhang2019,Shang2022}, in the $^3SD_1$ mixed pairing channel, the Cooper pair momentum adopts only these two orientations. As shown in Fig. \ref{fig:alpha-rho} (b), with increasing isospin asymmetry, $\rho_T$ becomes negative first, suggesting that the FFLO-ADG-P state emerges first. The transition from the ADG state to the FFLO-ADG-O state occurs where $\rho_T=0$ in the ADG state as shown in Fig. \ref{fig:alpha-rho} (d). Although the emergence of the FFLO-ADG-P state is expected to correspond to $\rho_L=0$ in the ADG state, whether it becomes the ground state can only be determined by direct comparison with the FFLO-ADG-O state. This also implies that the transition from the FFLO-ADG-O state to the FFLO-ADG-P state (O-P phase transition) is of first order. Compared with FFLO-AAG [Fig. \ref{fig:alpha-rho} (c)], the FFLO-ADG superfluid phase persists over a wider range of isospin asymmetry, demonstrating the result of the combined effects of the FFLO and ADG configurations. In addition, the angular dependence of the pairing gap significantly suppresses the normal-superfluid PS region, especially in the weak-coupling BCS regime at low asymmetry. At high density, the ADG and FFLO states can act in concert to eliminate PS entirely, although this effect gradually diminishes with decreasing density.

\begin{figure}[h]
            \centering
            \includegraphics[width=1.0\linewidth]{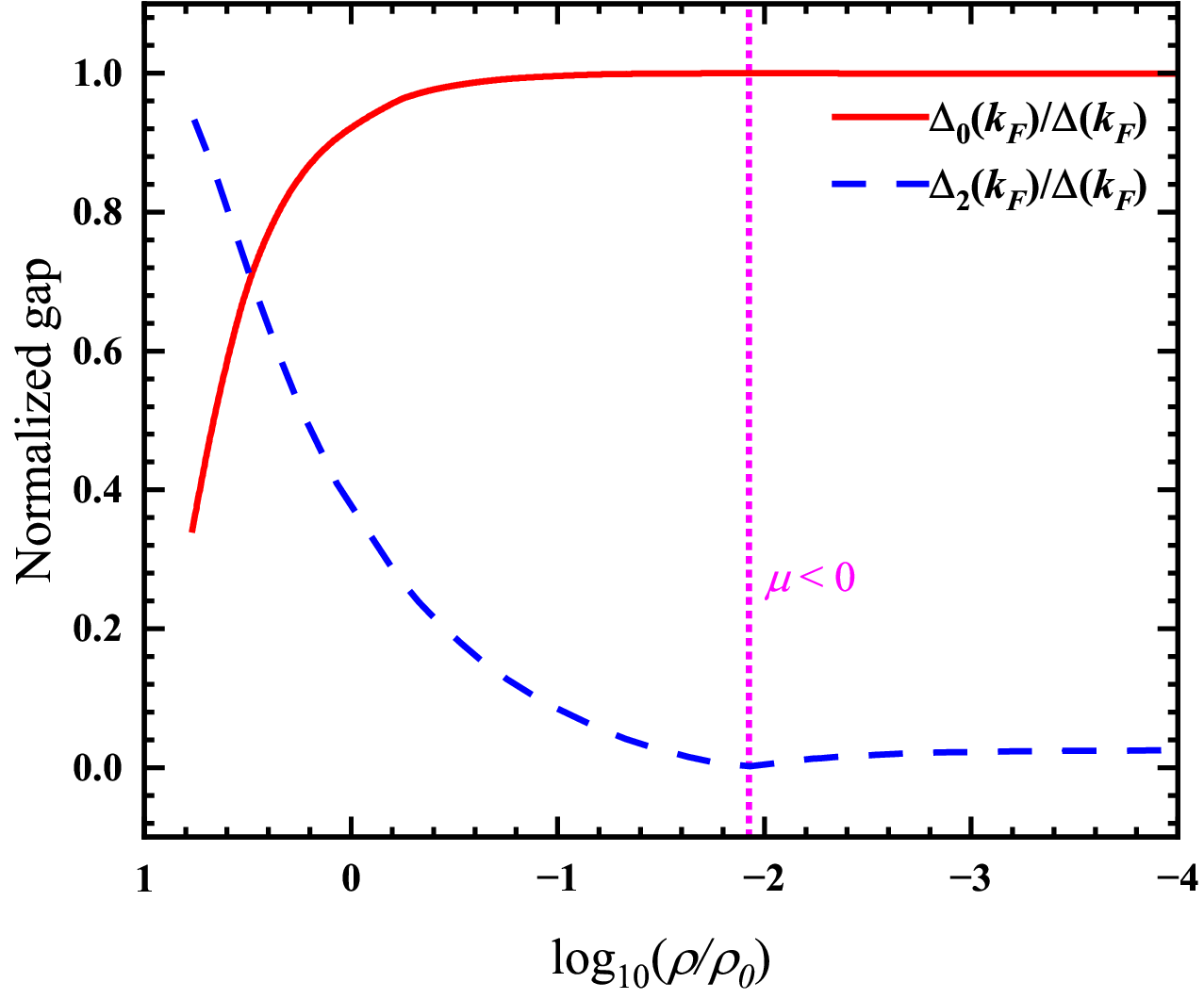}
            \caption{(Color online)
            Normalized gaps $\Delta_0(k_F)/\Delta(k_F)$ and $\Delta_2(k_F)/\Delta(k_F)$ as a function of number density at $T = 0.1$ MeV for symmetric nuclear matter. The thick red solid and blue dashed lines represent the gaps for the $^3S_1$ and $^3D_1$ channel, respectively. The thick magenta dotted lines identify the condition $\mu=0$.}
            \label{fig:wave}
        \end{figure}

 Of particular interest, near ($\alpha=0.3$, $\rho=2.5\rho_0$), a stable region of the ADG state is identified in Fig. \ref{fig:alpha-rho} (d) [see also the region with $\rho_T>0$ in Fig. \ref{fig:alpha-rho} (b)], which satisfies the criterion for a uniform superfluid phase. This feature clearly distinguishes $^3SD_1$ pairing from a conventional $S$-wave superfluid. The existence of this region is closely related to the contribution of the non-$S$-wave components. To elucidate this, Fig. \ref{fig:wave} presents the density dependence of the relative contributions of the $S$-wave and $D$-wave pairing gaps. One observes that the $D$-wave contribution decreases monotonically with decreasing density, whereas the $S$-wave component becomes increasingly dominant. Consequently, as the density decreases, the system evolves from being dominated by non-$S$-wave components to being $S$-wave dominated. More importantly, the systematic weakening of the ADG effects at lower densities originates not only from the reduced destructive influence of isospin asymmetry, but also from the diminishing fraction of non-$S$-wave components. At high density, for example around $\rho \sim 2.5\rho_0$, the system resembles a $D$-wave-dominated superfluid. As shown in \cite{Shang2022}, the presence of nodes and zero lines in the quasiparticle spectrum near the Fermi surface allows a sufficient number of excess neutrons to be accommodated, so that isospin asymmetry does not trigger instabilities in the superfluid density, nor the emergence of the FFLO state or normal–superfluid PS. 

At intermediate densities, such as $\rho \sim \rho_0$, the $D$-wave fraction decreases. Although a stable ADG phase does not exist as an isolated phase (since the FFLO state appears at a very small isospin asymmetry), the angular dependence of the pairing gap remains sufficient to eliminate the normal–superfluid PS in the low-asymmetry region \cite{Duan2025}. 
At lower densities, $\rho \sim 0.7\rho_0$, a qualitatively different behavior emerges, as illustrated in Fig. \ref{fig:alpha-rho} (d). In this regime, the ADG configuration can effectively suppress PS only at small isospin asymmetry, while PS reappears as the asymmetry increases. This reflects the significant reduction of the contribution of the $D$-wave, which limits the available phase space to accommodate excess neutrons and renders the ADG mechanism insufficient to stabilize the uniform superfluid at larger asymmetries. With further increasing asymmetry, the FFLO mechanism eventually suppresses PS. At even lower densities, the $D$-wave contribution becomes negligible, and the system approaches the behavior of a conventional $S$-wave superfluid. In general, Fig. \ref{fig:alpha-rho} (d) clearly illustrates a smooth evolution from a $D$-wave–dominated superfluid at high density to a $S$-wave superfluid at low density, accompanied by a corresponding weakening of the effects induced by the ADG.

        \begin{figure}[h]
            \centering
            \includegraphics[width=1.0\linewidth]{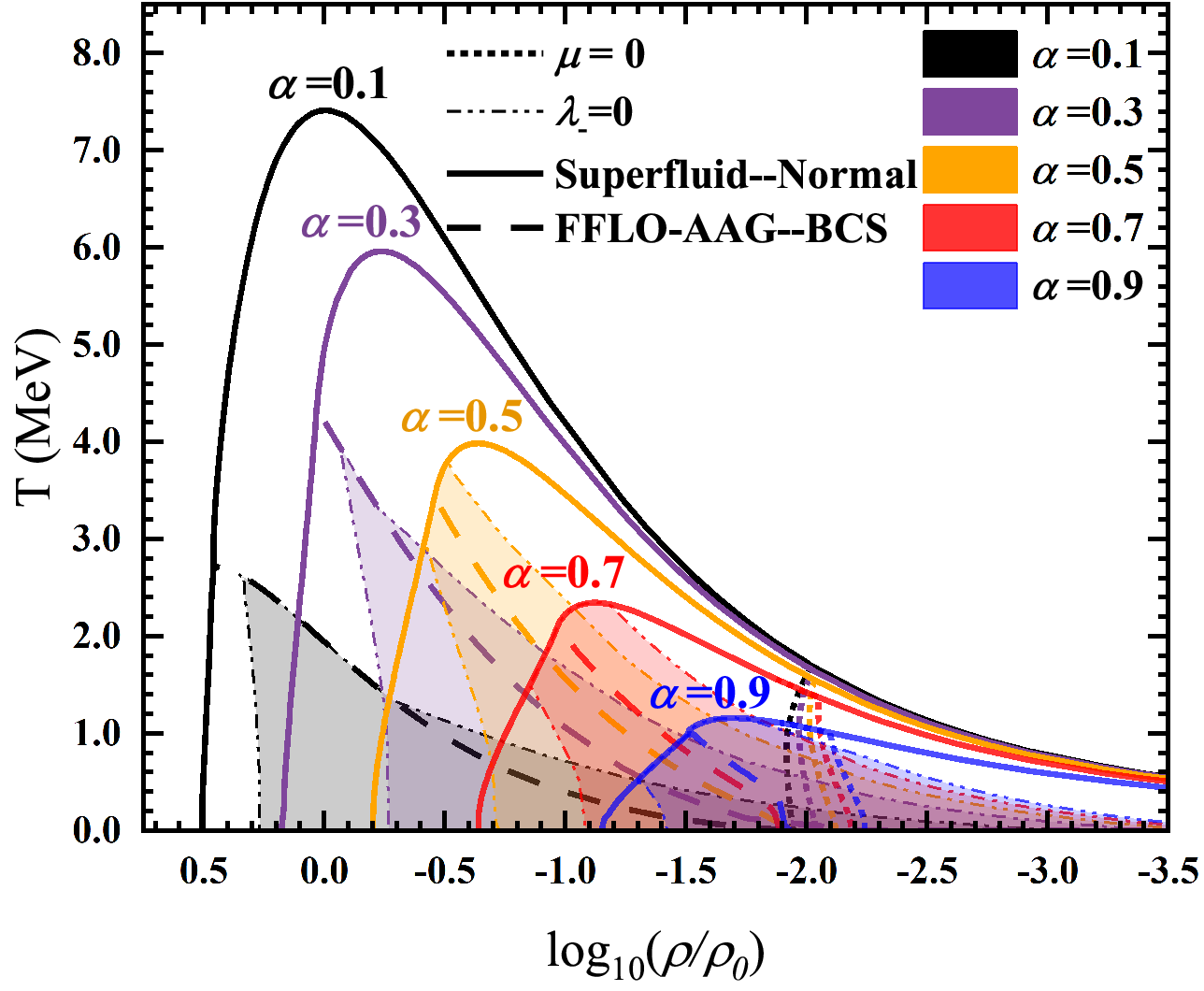}
            \caption{(Color online)
            Phase diagram of the FFLO-AAG state in the $T-\rho$ plane. The thick solid lines represent the transition from the superfluid to the normal states, while the thick dashed lines mark the transition from the FFLO-AAG state to the conventional BCS state. The thick dotted lines identify the condition $\mu=0$. The shaded regions represent the normal-superfluid PS domain, with the stability boundaries $\lambda_-$ marked by thin dash-dot-dotted lines. The black, purple, orange, red, and blue colors correspond to isospin asymmetries $\alpha=0.1$, $\alpha=0.3$, $\alpha=0.5$, $\alpha=0.7$, $\alpha=0.9$, respectively.}
            \label{fig:T-alpha-aag}
        \end{figure}

\begin{figure}[h]
            \centering
            \includegraphics[width=1.0\linewidth]{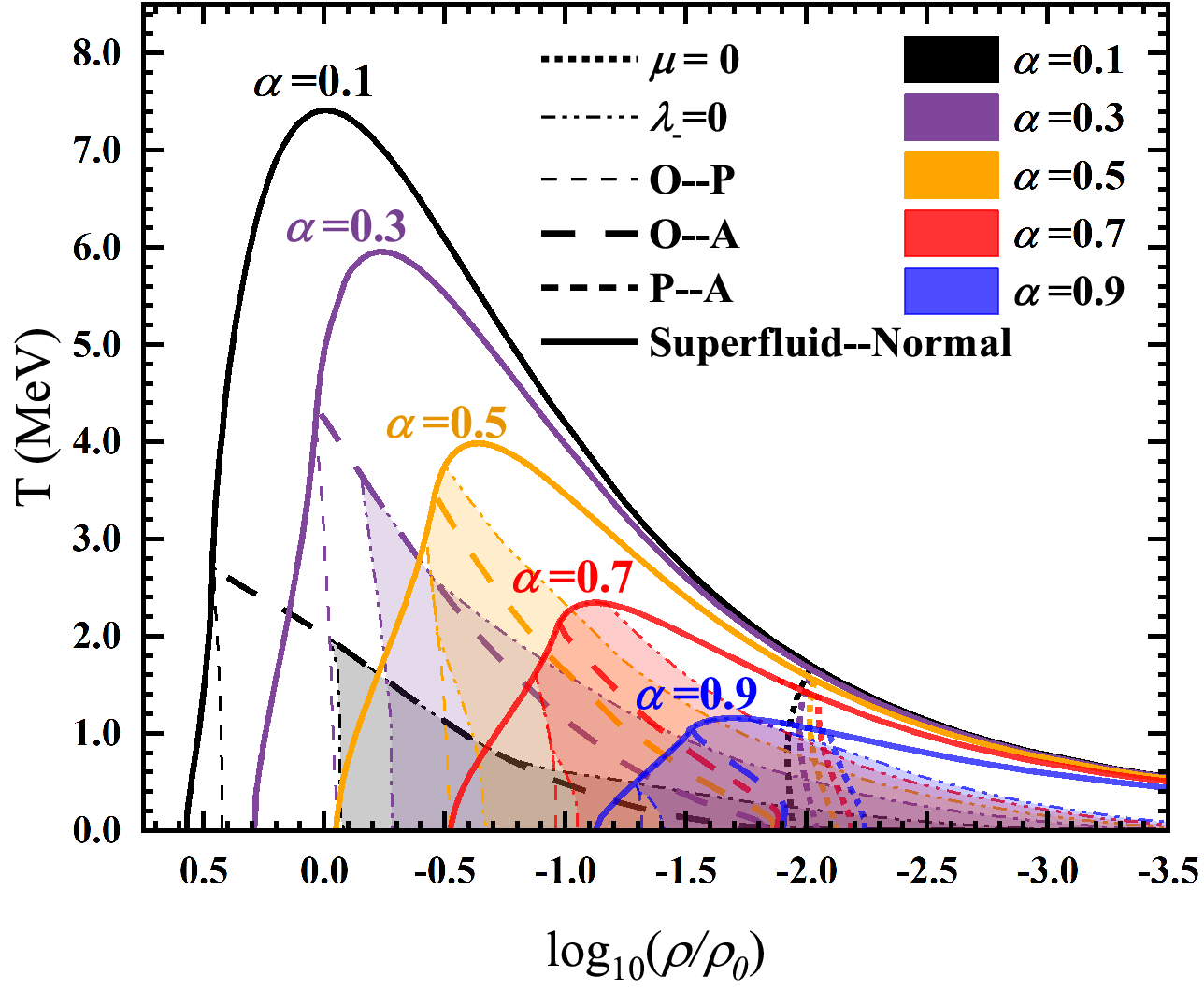}
            \caption{(Color online)
            Same as Fig. \ref{fig:T-alpha-aag}, but for the FFLO-ADG state. The FFLO-ADG state splits into the FFLO-ADG-O and FFLO-ADG-P states. The thin dashed, thick dashed, and thick short-dashed lines denote the transition from FFLO-ADG-O to FFLO-ADG-P (O-P), from FFLO-ADG-O to ADG (O-A), and from FFLO-ADG-P to ADG (P-A), respectively.
            }
            \label{fig:T-alpha-adg}
        \end{figure}

        To further characterize the phase diagram of the BCS-BEC crossover with FFLO state, Figs. \ref{fig:T-alpha-aag} and \ref{fig:T-alpha-adg} display the phase diagram in the $T-\rho$ plane for different asymmetries for the FFLO-AAG and FFLO-ADG states, respectively. The phase diagram shown in Fig. \ref{fig:T-alpha-aag} closely resembles that of $S$-wave pairing \cite{Stein2012}, since the angle-averaged gap is equivalent to an $S$-wave gap. For fixed asymmetry and density, as the temperature increases, thermal smearing of the Fermi surfaces becomes more pronounced, gradually weakening the effectiveness of the Cooper pair momentum in the FFLO state to compensate for the Fermi-surface mismatch. Consequently, the system undergoes a second-order phase transition from the FFLO state to the conventional BCS state. With increasing isospin asymmetry, the role of the Cooper pair momentum becomes more significant. As shown in Fig. \ref{fig:T-alpha-aag}, the relative region in which the FFLO state exists expands. Nevertheless, as indicated in Fig. \ref{fig:alpha-rho}, the Cooper pair momentum does not play a significant role in the strong-coupling BEC regime, and the FFLO state is completely absent in this region. Furthermore, the critical temperature $T_c$, corresponding to the disappearance of the pairing gap, rapidly decreases with increasing asymmetry in the weak-coupling BCS regime. However, this suppression is much less pronounced in the strong-coupling BEC regime, where stronger interactions provide robust resistance to the pair-breaking effects of the asymmetry. 
        
        At high densities, the normal-superfluid PS region appears only in the FFLO state. With further decreasing density, a PS region also emerges in the conventional BCS state. Moreover, as the isospin asymmetry increases, for example, at $\alpha=0.5$, the homogeneous FFLO and homogeneous BCS states may be separated by an inhomogeneous mixed phase. It is also worth noting that the relative extent of the homogeneous superfluid phase is significantly reduced with increasing asymmetry. In addition, in the strong-coupling BEC regime at low temperatures, a mixed phase composed of superfluid and normal components with spatially inhomogeneous asymmetries is expected to occur in asymmetric nuclear matter. This PS-BEC phase is consistent with that reported in Ref. \cite{Stein2012}.

        Fig. \ref{fig:T-alpha-adg} presents the phase diagram for the FFLO-ADG state. For a fixed asymmetry, the FFLO-ADG-P and FFLO-ADG-O states are located in the relative high- and low-density regions, respectively, and are separated by a first-order phase transition. Beyond lifting the orientational degeneracy of the FFLO state, the combined mechanisms of angle-dependent pairing and finite Cooper momentum pairing significantly suppress the normal-superfluid PS region compared to the FFLO-AAG case. In addition, the PS region is nearly absent in the FFLO-ADG-P state and occurs predominantly in the FFLO-ADG-O state. However, these two effects arising from the angle-dependent pairing gap are present only within the region where the FFLO state exists and are completely suppressed in the BEC regime.

    \section{Summary and outlook}\label{summary and outlook}
    Summarily, we have systematically investigated the phase structure of the BCS-BEC crossover for the $np$ superfluid in asymmetric nuclear matter, with particular emphasis on the role of the angular dependence of the pairing gap and the emergence of FFLO states. By constructing phase diagrams in the $T-\alpha$, $\alpha-\rho$, $T-\rho$ planes for both angle-averaged and angle-dependent gap scenarios, we systematically elucidate the roles played by the FFLO and ADG mechanisms, together with normal-superfluid PS, in the resulting phase structure.

    The results confirm the picture that the BCS-BEC crossover in asymmetric nuclear matter is primarily density-driven, reflecting the effective enhancement of the coupling strength of the $np$ pairs with decreasing density. In the weak-coupling BCS regime, isospin asymmetry strongly suppresses the stability of the homogeneous superfluid phase and tends to drive the system toward normal-superfluid PS; nevertheless, the FFLO and ADG mechanisms can, to some extent, alleviate the deleterious effects of asymmetry on superfluidity. In contrast, in the BEC regime, the FFLO and ADG mechanisms are completely excluded, while phase separation persists, providing a possible way to accommodate the excess neutrons. In particular, in the low-temperature and high-asymmetry region of the BEC regime, the system favors an inhomogeneous mixed phase in which the superfluid component consists of a BEC of deuterons. Throughout the entire phase diagram, the effects of the FFLO mechanism, the angular dependence of the pairing gap, and phase separation are ultimately quenched by thermal excitations as the temperature increases.

By comparing the AAG and ADG treatments, we find that the angular dependence of the pairing gap itself does not extend the asymmetry window in which the pairing survives. However, in combination with the FFLO state, it can enlarge the asymmetry range where the superfluid phase remains stable and significantly suppresses the normal-superfluid PS in the BCS regime, as quantitatively reflected in the stability conditions against PS in the homogeneous superfluid state. Notably, at high density, the combined effects of FFLO and ADG can nearly eliminate all PS regions. As density decreases, the influence of the angular dependence gradually diminishes, partly because the effective $np$ coupling strengthens, reducing the destructive effect of asymmetry on $np$ pairing and thereby weakening the ADG-induced suppression. More importantly, the fraction of the $D$-wave component in the $^3SD_1$ channel decreases with density, directly reducing the angular dependence of the pairing gap. In general, the phase diagram exhibits a smooth evolution from a $D$-wave–dominated superfluid at high density to a conventional $S$-wave superfluid at low density, accompanied by a systematic attenuation of ADG-induced effects. In the weak-coupling regime, the angular dependence also leads to a splitting of the superfluid density into transverse and longitudinal components, which in turn splits the FFLO state into two distinct ADG states, FFLO-ADG-O and FFLO-ADG-P, corresponding to the nondegenerate orientations of the Cooper pair momentum orthogonal and perpendicular to the axial symmetry of the ADG state. These two states occupy different density regions and are separated by a first-order phase transition.

    
    In the present work, the bare $NN$ potential and a free single-particle spectrum are adopted with the medium effects such as screening and self-energy corrections neglected. Although this approach is sufficient to capture the qualitative features of the phase diagram and the interplay between the ADG and FFLO mechanisms, as well as the normal-superfluid PS, incorporating medium effects will be essential for a more quantitative determination of phase boundaries and stability conditions. In addition, at low densities, dilute nuclear matter is also expected to exhibit clusters with higher mass numbers, notably $^3$He, $^3$H, and $^4$He \cite{Heckel2009,Typel2010}, which may coexist in statistical equilibrium with the superfluid and normal components. 
    Such clustering effects can qualitatively modify the phase structure, particularly in the BEC regime, and call for a unified treatment of pairing correlations and cluster formation. 

    \section*{Acknowledgments}
    This work was supported by CAS Project for Young Scientists in Basic Research YSBR-088; the National Natural Science Foundation of China under Grant No. 12375117; the Youth Innovation Promotion Association of Chinese Academy of Sciences (Grant No. Y2021414).
    \bibliography{text}
\end{document}